\documentclass[amsmath,amssymb,11pt]{article}
\usepackage{amsmath,amsfonts,latexsym,graphicx,amssymb}
\date{}
\usepackage{amsfonts}
\usepackage{bm}

\newcommand{\ot}{{\,\otimes\,}}
\newcommand{{\Cd}}{{\mathbb{C}^d}}
\newcommand{\sbsigma}{{\mbox{\scriptsize \boldmath $\sigma$}}}

\newcommand{\bsigma}{{\mbox{ \boldmath $\sigma$}}}

\def\oper{{\mathchoice{\rm 1\mskip-4mu l}{\rm 1\mskip-4mu l}%
{\rm 1\mskip-4.5mu l}{\rm 1\mskip-5mu l}}}
\def\<{\langle}
\def\>{\rangle}

\begin{document}

\title{\textbf{How to construct entanglement witnesses}}
\author{Dariusz Chru\'sci\'nski and Andrzej
Kossakowski \\
Institute of Physics, Nicolaus Copernicus University,\\
Grudzi\c{a}dzka 5/7, 87--100 Toru\'n, Poland}

\maketitle

\begin{abstract}
We present very simple method for constructing indecomposable
entanglement witnesses out of a given pair --- an entanglement
witness $W$ and the corresponding state detected by $W$. This
method may be used to produce new classes of atomic witnesses
which are able to detect the `weakest' quantum entanglement.
Actually, it works perfectly in the multipartite case, too.
Moreover, this method provides a powerful tool for constructing
new examples of bound entangled states.
\end{abstract}

\section{Introduction}

One of the most important problems of quantum information theory
\cite{QIT,Horodecki-review} is the characterization of mixed
states of composed quantum systems. In particular it is of primary
importance to test whether a given quantum state  is separable or
entangled. For low dimensional systems there exists simple
necessary and sufficient condition for separability. The
celebrated Peres-Horodecki criterium \cite{Peres,Pawel} states
that a state of a bipartite system living in $\mathbb{C}^2 \ot
\mathbb{C}^2$ or $\mathbb{C}^2 \ot \mathbb{C}^3$ is separable iff
its partial transpose is positive, i.e. a state is PPT.
Unfortunately, for higher-dimensional systems there is no single
universal separability condition.

The most general approach to characterize quantum entanglement
uses a notion of an entanglement witness (EW)
\cite{HHH,Terhal1,Terhal2}. A Hermitian operator $W$ defined on a
tensor product $\mathcal{H}=\mathcal{H}_1 \ot \mathcal{H}_2$ is
called  an EW iff 1) $\mbox{Tr}(W\sigma_{\rm sep})\geq 0$ for all
separable states $\sigma_{\rm sep}$, and 2) there exists an
entangled state $\rho$ such that $\mbox{Tr}(W\rho)<0$ (one says
that $\rho$ is detected by $W$). It turns out that a state is
entangled if and only if it is detected by some EW \cite{HHH}.
There was a considerable effort in constructing and analyzing the
structure of EWs [6--15]. There were also attempts for their
experimental realizations \cite{EX,Wu} and several procedures for
optimizing EWs for arbitrary states were proposed
\cite{O,O1,O2,O3}.

The simplest way to construct EW is to define $W = P + (\oper \ot
\tau)Q$, where $P$ and $Q$ are positive operators, and  $(\oper
\ot \tau)Q$ denotes partial transposition. It is easy to see that
$\mbox{Tr}(W\sigma_{\rm sep})\geq 0$ for all separable states
$\sigma_{\rm sep}$, and hence if $W$ is non-positive, then it is
EW. Such EWs are said to be decomposable \cite{Lew}. Note,
however, that decomposable EW cannot detect PPT entangled state
(PPTES) and, therefore, such EWs are useless in the search for
bound entangled state. Unfortunately,  there is no general method
to construct indecomposable EW and only very few examples of
indecomposable EWs are available in the literature.

In the present paper we propose very simple method for
constructing indecomposable EWs. If we are given one
indecomposable EW $W_0$ and the corresponding state $\rho_0$
detected by $W_0$, then we are able to construct an open convex
set of indecomposable EWs detecting $\rho_0$, and an open convex
set of PPTES detected by $W_0$. Hence, out of a given pair
$(W_0,\rho_0)$ we construct huge classes of new EWs and PPTES,
respectively. In particular, we may apply this method to construct
so called atomic EWs which are able to detect the `weakest'
quantum entanglement (i.e. PPTES $\rho$ such that both Schmidt
number \cite{SN} of $\rho$ and its partial transposition $(\oper
\ot \tau)\rho$ does not exceed 2). We stress that  proposed method
is very general and it works perfectly for multipartite case.

The paper is organized as follows: in the next section we
introduce a natural hierarchy of convex cones in the space of EWs.
This hierarchy explains the importance of indecomposable and
atomic EWs. Section \ref{PPT}  presents our method for
constructing indecomposable EWs. Section \ref{ATOM} provides
construction of atomic EWs and it is illustrated by a new class of
such witnesses. Finally, in section \ref{MULTI} we generalize our
construction for multipartite case.  A brief discussion is
included in the last section.

\section{A hierarchy of entanglement witnesses}

Consider a space $\mathcal{P}$ of positive operators in
$\mathcal{B}(\mathcal{H}_1 \ot \mathcal{H}_2)$. There is a natural
family of convex cones in $\mathcal{P}$:
\begin{equation}\label{}
    \mathbf{V}_r = \{\, \rho \in \mathcal{P}\ |\
    \mathrm{SN}(\rho) \leq r\, \}  \ ,
\end{equation}
where $\mathrm{SN}(\rho)$ denotes the Schmidt number of
(unnormalized) positive operator $\rho$ \cite{SN}. One has the
following chain of inclusions
\begin{equation}\label{V-k}
\mathbf{V}_1  \subset \ldots \subset  \mathbf{V}_d = \mathcal{P}\
,
\end{equation}
where $d = \min\{d_1,d_2\}$, and $d_k = \mbox{dim}\,
\mathcal{H}_k$.  Clearly, $\mathbf{V}_1$ is a cone of separable
(unnormalized) states and $\mathbf{V}_d \smallsetminus
\mathbf{V}_1$ stands for a set of entangled states. Note, that a
partial transposition $(\oper \ot \tau)$ gives rise to another
family of cones:
\begin{equation}\label{}
     \mathbf{V}^l = (\oper \ot \tau)\mathbf{V}_l \ ,
\end{equation}
such that $ \mathbf{V}^1  \subset \ldots \subset  \mathbf{V}^d$.
 One has
$\mathbf{V}_1 = \mathbf{V}^1$, together with the following
hierarchy of inclusions:
\begin{equation}\label{}
    \mathbf{V}_1 = \mathbf{V}_1 \cap \mathbf{V}^1 \subset \mathbf{V}_2 \cap
    \mathbf{V}^2   \subset \ldots \subset \mathbf{V}_d \cap \mathbf{V}^d \
    .
\end{equation}
Note, that $\mathbf{V}_d \cap \mathbf{V}^d$ is a convex set of PPT
(unnormalized) states. Finally, $\mathbf{V}_r \cap \mathbf{V}^s$
is a convex subset of PPT states $\rho$  such that
$\mathrm{SN}(\rho) \leq r$ and $\mathrm{SN}[(\oper \ot \tau)\rho]
\leq s$.

Now, in the set of entanglement witnesses $\mathbf{W}$ one may
introduce the  family of dual cones:
\begin{equation}\label{}
\mathbf{W}_r = \{\, W\in \mathcal{B}(\mathcal{H}_1 \ot
\mathcal{H}_2) |\ \mathrm{Tr}(W\rho) \geq 0\ , \ \rho \in
\mathbf{V}_r\, \}\ .
\end{equation}
One has
\begin{equation}\label{}
\mathcal{P} = \mathbf{W}_d  \subset \ldots \subset \mathbf{W}_1 \
.
\end{equation}
Clearly, $\mathbf{W} = \mathbf{W}_1 \smallsetminus \mathbf{W}_d$.
Moreover, for any $k>l$, entanglement witnesses from $\mathbf{W}_l
\smallsetminus \mathbf{W}_k$  can detect entangled states from
$\mathbf{V}_k \smallsetminus  \mathbf{V}_l$, i.e. states $\rho$
with Schmidt number $l < \mathrm{SN}(\rho) \leq k$. In particular
$W \in \mathbf{W}_k \smallsetminus \mathbf{W}_{k+1}$ can detect
state $\rho$ with $\mathrm{SN}(\rho)=k$.

Finally, let us consider the following class
\begin{equation}\label{}
    \mathbf{W}_r^s = \mathbf{W}_r + (\oper \ot \tau)\mathbf{W}_s\
    ,
\end{equation}
that is, $W \in \mathbf{W}_r^s$ iff
\begin{equation}\label{}
    W = P + (\oper \ot \tau)Q\ ,
\end{equation}
with $P \in \mathbf{W}_r$ and $Q \in \mathbf{W}_s$. Note, that
$\mathrm{Tr}(W\rho) \geq 0$ for all $\rho \in \mathbf{V}_r \cap
\mathbf{V}^s$. Hence such $W$ can detect PPT states $\rho$ such
that $\mathrm{SN}(\rho) \geq r$ and $\mathrm{SN}[(\oper \ot
\tau)\rho] \geq s$. Entanglement witnesses from $\mathbf{W}_d^d$
are called decomposable \cite{Lew}. They cannot detect PPT states.
One has the following chain of inclusions:
\begin{equation}\label{}
    \mathbf{W}_d^d\, \subset\, \ldots\, \subset\, \mathbf{W}^2_2\, \subset\, \mathbf{W}^1_1\,
    \equiv\,    \mathbf{W}\ .
\end{equation}
To deal with PPT states one needs indecomposable witnesses from
$\mathbf{W}^{\rm ind} := \mathbf{W} \smallsetminus
\mathbf{W}_d^d$. The `weakest' entanglement can be detected by
elements from $\mathbf{W}^{\rm atom} := \mathbf{W} \smallsetminus
\mathbf{W}_2^2$. We shall call them {\em atomic entanglement
witnesses}. It is clear that $W$ is an atomic entanglement witness
if there is an entangled state $\rho \in \mathbf{V}_2 \cap
\mathbf{V}^2$ such that $\mathrm{Tr}(W \rho) <0$. The knowledge of
atomic witnesses, or equivalently atomic maps,  is crucial:
knowing this set we would be able to distinguish all entangled
states from separable ones.

\section{Detecting PPT entangled states} \label{PPT}

Suppose that a PPT entangled state $\rho_0$ in $\mathcal{H}_1 \ot
\mathcal{H}_2$ is detected by an entanglement witness $W_0$, that
is
\begin{equation}\label{}
    \mbox{Tr}(W_0\rho_0)< 0\ .
\end{equation}
It is clear that in the vicinity of $\rho_0$ there are other PPT
entangled states  detected by the same witness $W_0$.  Let
$\sigma_{\rm sep}$ be an arbitrary separable state and consider
the following convex combination
\begin{equation}\label{cc}
    \rho_\alpha = (1-\alpha)\rho_0 + \alpha \sigma_{\rm sep} \ .
\end{equation}
It is evident that $\rho_\alpha$ is PPT for any $\alpha \in
[0,1]$. Moreover, for any $0 \leq \alpha <
\alpha_{[\rho_0,\sigma_{\rm sep}]}$, with
\begin{equation}\label{}
\alpha_{[\rho_0,\sigma_{\rm sep}]} := \sup\, \{ \alpha \in [0,1]\
|\ \mbox{Tr}(W_0\rho_\alpha)< 0 \} \ ,
\end{equation}
$\rho_\alpha$ is entangled. This construction gives rise to an
open convex set
\begin{equation}\label{}
\mathcal{S}^{\rm PPT}[W_0|\rho_0] := \Big\{ \rho_\alpha\ \Big| \ 0
\leq \alpha < \alpha_{[\rho_0,\sigma_{\rm sep}]}\ \ \& \ {\rm
aribitrary}\ \ \sigma_{\rm sep} \ \Big\} \ .
\end{equation}
All elements from $\mathcal{S}^{\rm PPT}[W_0|\rho_0]$ are PPT
entangled states detected by $W_0$. On the other hand in the
vicinity of $W_0$ there are other entanglement witnesses detecting
our original PPT state $\rho_0$. Indeed, let
 $P$ be an arbitrary positive semidefinite operator in $\mathcal{B}(\mathcal{H}_1 \ot
 \mathcal{H}_2)$ and consider one-parameter family of operators
\begin{equation}\label{W-lambda}
    W_\lambda = W_0 + \lambda P\ , \ \ \ \ \ \lambda \geq 0 \ .
\end{equation}
Let us observe that for any $0 \leq \lambda < \lambda_{[W_0,P]}$
with
\begin{equation}\label{}
\lambda_{[W_0,P]} := \sup\, \{ \lambda \geq 0 \ |\
\mbox{Tr}(W_\lambda\rho_0)< 0 \} \ ,
\end{equation}
$W_\lambda$ is an indecomposable EW detecting a PPT state
$\rho_0$. This construction gives rise to a dual open convex set
\begin{equation}\label{}
\mathcal{W}^{\rm ind}[W_0|\rho_0] := \Big\{ \ W_\lambda\ \Big| \ 0
\leq \lambda < \lambda_{[W_0,P]}\ \ \&\ \ {\rm aribitrary }\ P\geq
0 \ \Big\} \ .
\end{equation}
Summarizing, having a pair of a PPTES $\rho_0$ and an
indecomposable EW $W_0$ we may construct two open convex sets:
$\mathcal{S}^{\rm PPT}[W_0|\rho_0]$ containing PPTES detected by
$W_0$ and $\mathcal{W}^{\rm ind}[W_0|\rho_0]$ containing
indecomposable EW detecting $\rho_0$. It shows that for any
$\rho_1, \rho_2 \in \mathcal{S}^{\rm PPT}[W_0|\rho_0]$ any convex
combination
\begin{equation}\label{}
p_1 \rho_1 +  p_2 \rho_2\, \in\, \mathcal{S}^{\rm
PPT}[W_0|\rho_0]\ ,
\end{equation}
and hence defines a PPTES. Similarly, for any $W_1, W_2 \in
\mathcal{W}^{\rm ind}[W_0|\rho_0]$ any convex combination
\begin{equation}\label{}
w_1 W_1 +  w_2 W_2\, \in\, \mathcal{W}^{\rm ind}[W_0|\rho_0]\ ,
\end{equation}
and hence defines a indecomposable EW. Therefore, the above
constructions provide a methods to produce new PPTES and new
indecomposable EW out of a single pair $(\rho_0,W_0)$.

Note, that this construction may be easily continued. Let us take
an arbitrary EW $W'$ from $\mathcal{W}^{\rm ind}[W_0|\rho_0]$
(different from $W_0$). It is easy to find PPTES from
$\mathcal{S}^{\rm PPT}[W_0|\rho_0]$ detected by $W'$: indeed, any
state in $\mathcal{S}^{\rm PPT}[W_0|\rho_0]$ has a form (\ref{cc})
and hence
\begin{equation}\label{}
    \mbox{Tr}(W' \rho_\alpha) = (1-\alpha)\mbox{Tr}(W'\rho_0) +
    \alpha\mbox{Tr}(W' \sigma_{\rm sep})\ .
\end{equation}
Therefore, one has $\mbox{Tr}(W' \rho_\alpha)<0$ for
\begin{equation}\label{a-new}
    \alpha < \frac{-\mbox{Tr}(W'\rho_0)}{-\mbox{Tr}(W'\rho_0) + \mbox{Tr}(W'\sigma_{\rm
    sep})} \leq 1\ .
\end{equation}
Now,  $W'$ and $\rho'=\rho_\alpha$ with $\alpha$ satisfying
(\ref{a-new}) defines a new pair which may be used as a starting
point for the  construction of $\mathcal{S}^{\rm PPT}[W'|\rho']$
and $\mathcal{W}^{\rm ind}[W'|\rho']$.

\section{Constructing atomic entanglement witnesses} \label{ATOM}

Suppose now, that we are given a `weakly entangled' PPTES, i.e. a
state $\rho_0 \in \mathbf{V}_2 \cap \mathbf{V}^2$ and let $W_0$ be
the corresponding atomic EW. Following our construction we define
\begin{equation}\label{}
\mathcal{S}^{2}_2[W_0|\rho_0] \subset \mathbf{V}_2 \cap
\mathbf{V}^2 \ ,
\end{equation}
such that each element from $\mathcal{S}^{2}_2[W_0|\rho_0]$ is
detected by the same witness $W_0$. Similarly, we define a set of
atomic witnesses
\begin{equation}\label{}
\mathcal{W}^{\rm atom}[W_0|\rho_0] \subset \mathbf{W}^{\rm atom} \
,
\end{equation}
such that each element from $\mathcal{W}^{\rm atom}[W_0|\rho_0]$
detects our original state $\rho_0$. Both sets
$\mathcal{S}^{2}_2[W_0|\rho_0]$ and $\mathcal{W}^{\rm
atom}[W_0|\rho_0]$ are open and convex.

Note, that knowing atomic EWs one may detect all entangled states.
Moreover, it was conjectured by Osaka \cite{Osaka} that all EWs in
$\mathcal{B}(\mathbb{C}^3 \ot \mathbb{C}^3)$ may be represented as
a sum of decomposable and atomic witnesses. To the best of our
knowledge this conjecture is still open. It shows that the
knowledge of atomic EWs is crucial both from physical and purely
mathematical point of view. Let us illustrate the construction of
atomic EWs by the following

\noindent {\bf Example}: new class of atomic EWs in $3\ot 3$.

It is well known that there is a direct relation between
entanglement witnesses in $\mathcal{B}(\mathcal{H}_1 \ot
\mathcal{H}_2)$ and positive maps $\Lambda :
\mathcal{B}(\mathcal{H}_1)
\longrightarrow\mathcal{B}(\mathcal{H}_2)$. Due to the
Choi-Jamio{\l}kowski isomorphism \cite{J,C} one has
\begin{equation}\label{JC}
    \varphi \ \longrightarrow \ W_\varphi := \sum_{i,j=1}^{d_1}
    e_{ij} \ot \varphi(e_{ij})\ ,
\end{equation}
with $d_1 = \mbox{dim}\,\mathcal{H}_1$. In what follows we are
using the following notation: $(e_1,\ldots,e_d)$ denotes an
orthonormal basis in $\mathbb{C}^d$, and $e_{ij} = |e_i\>\<e_j|$.
 Consider now the following
operator in $M_3 \ot M_3$ which is related via
Choi-Jamio{\l}kowski isomorphism to the celebrated Choi map
\cite{C}\footnote{The (unnormalized) Choi map $\varphi : M_3
\longrightarrow M_3$ is defined as follows:
\begin{eqnarray*}\label{}
    \varphi(e_{11}) = e_{11} + e_{22} \ , \ \ \
    \varphi(e_{22}) = e_{22} + e_{33} \ , \ \ \
    \varphi(e_{33}) = e_{33} + e_{11} \ ,
\end{eqnarray*}
and $\varphi(e_{ij}) = - e_{ij}$, for $i \neq j$. }
\begin{equation}\label{W0-C}
 W_0\ = \  \left( \begin{array}{ccc|ccc|ccc}
 1 &  \cdot& \cdot& \cdot& -1 & \cdot& \cdot& \cdot & -1 \\
 \cdot& 1 &\cdot& \cdot& \cdot& \cdot& \cdot& \cdot& \cdot\\
 \cdot& \cdot& \cdot & \cdot& \cdot& \cdot& \cdot& \cdot& \cdot  \\ \hline
 \cdot& \cdot& \cdot& \cdot & \cdot& \cdot& \cdot& \cdot& \cdot \\
 -1 & \cdot& \cdot& \cdot& 1 & \cdot& \cdot& \cdot& -1 \\
 \cdot& \cdot& \cdot& \cdot& \cdot& 1& \cdot & \cdot& \cdot  \\ \hline
 \cdot& \cdot& \cdot& \cdot& \cdot& \cdot & 1& \cdot& \cdot \\
 \cdot & \cdot& \cdot& \cdot& \cdot& \cdot& \cdot& \cdot& \cdot \\
 -1& \cdot& \cdot& \cdot& -1 & \cdot& \cdot& \cdot& 1
  \end{array} \right)\ ,
\end{equation}
where to maintain more transparent form we replace all zeros by
dots. It was shown by Ha \cite{Ha} that $W_0$ is atomic. The proof
is based on the construction of a state in $\mathbf{V}_2 \cap
\mathbf{V}^2$ detected by $W_0$. Actually, Ha constructed a whole
one-parameter family of such states. For any $0 < \gamma < 1$ let
us define
\begin{equation}
  \rho_\gamma\ = \  \frac{1}{N_\gamma}\left( \begin{array}{ccc|ccc|ccc}
 1 &  \cdot& \cdot& \cdot& 1 & \cdot& \cdot& \cdot & 1 \\
 \cdot& a_\gamma &\cdot& \cdot& \cdot& \cdot& \cdot& \cdot& \cdot\\
 \cdot& \cdot& b_\gamma & \cdot& \cdot& \cdot& \cdot& \cdot& \cdot  \\ \hline
 \cdot& \cdot& \cdot& b_\gamma & \cdot& \cdot& \cdot& \cdot& \cdot \\
 1 & \cdot& \cdot& \cdot& 1 & \cdot& \cdot& \cdot& 1 \\
 \cdot& \cdot& \cdot& \cdot& \cdot& a_\gamma& \cdot & \cdot& \cdot  \\ \hline
 \cdot& \cdot& \cdot& \cdot& \cdot& \cdot & a_\gamma& \cdot& \cdot \\
 \cdot & \cdot& \cdot& \cdot& \cdot& \cdot& \cdot& b_\gamma& \cdot \\
 1& \cdot& \cdot& \cdot& 1 & \cdot& \cdot& \cdot& 1
  \end{array} \right)\ ,
\end{equation}
with
\begin{equation}\label{}
    a_\gamma = \frac 13 (\gamma^2 + 2) \ , \ \ \ b_\gamma
    = \frac 13 (\gamma^{-2} + 2)\ ,
\end{equation}
and the normalization factor
\begin{equation}\label{N-3}
    N_\gamma = 7 + \gamma^2 + \gamma^{-2}\ .
\end{equation}
It was shown \cite{Ha} that $\rho_\gamma \in \mathbf{V}_2 \cap
\mathbf{V}^2$ and $\mbox{Tr}(W_0 \rho_\gamma) =
(\gamma^2-1)/N_\gamma$. Hence, for $\gamma < 1$ the state
$\rho_\gamma$ is entangled (and $W_0$ is indecomposable
EW).\footnote{Actually, for $\gamma=1$ one has
\begin{equation}
  \rho_{\gamma=1}\ = \  \frac{1}{9}\left( \begin{array}{ccc|ccc|ccc}
 1 &  \cdot& \cdot& \cdot& 1 & \cdot& \cdot& \cdot & 1 \\
 \cdot& 1 &\cdot& \cdot& \cdot& \cdot& \cdot& \cdot& \cdot\\
 \cdot& \cdot& 1 & \cdot& \cdot& \cdot& \cdot& \cdot& \cdot  \\ \hline
 \cdot& \cdot& \cdot& 1 & \cdot& \cdot& \cdot& \cdot& \cdot \\
 1 & \cdot& \cdot& \cdot& 1 & \cdot& \cdot& \cdot& 1 \\
 \cdot& \cdot& \cdot& \cdot& \cdot& 1 & \cdot & \cdot& \cdot  \\ \hline
 \cdot& \cdot& \cdot& \cdot& \cdot& \cdot & 1 & \cdot& \cdot \\
 \cdot & \cdot& \cdot& \cdot& \cdot& \cdot& \cdot& 1 & \cdot \\
 1& \cdot& \cdot& \cdot& 1 & \cdot& \cdot& \cdot& 1
  \end{array} \right)\ ,
\end{equation}
and it is known \cite{Pawel} that this state is separable.}
 It
is therefore clear that if $\gamma_1,\ldots,\gamma_K \in (0,1)$,
then any convex combination
\begin{equation}\label{}
    p_1 \rho_{\gamma_1} + \ldots + p_K \rho_{\gamma_K}
\end{equation}
defines an entangled state in $\mathbf{V}_2 \cap \mathbf{V}^2$
detected by $W_0$.

Consider now the following maximally entangled state in
$\mathbb{C}^3 \ot \mathbb{C}^3$:
\begin{equation}\label{}
    \psi = \frac{1}{\sqrt{3}} ( e_1 \ot e_3 + e_2 \ot e_1 + e_3 \ot
    e_2) \ ,
\end{equation}
and let $P = 3|\psi\>\<\psi|$. Define $W_\lambda = W_0 + \lambda
P$. It is given by the following matrix
\begin{equation}\label{}
 W_\lambda\ = \  \left( \begin{array}{ccc|ccc|ccc}
 1 &  \cdot& \cdot& \cdot& -1 & \cdot& \cdot& \cdot & -1 \\
 \cdot& 1 &\cdot& \cdot& \cdot& \cdot& \cdot& \cdot& \cdot\\
 \cdot& \cdot& \lambda & \lambda & \cdot& \cdot& \cdot& \lambda & \cdot  \\ \hline
 \cdot& \cdot& \lambda & \lambda & \cdot& \cdot& \cdot& \lambda & \cdot \\
 -1 & \cdot& \cdot& \cdot& 1 & \cdot& \cdot& \cdot& -1 \\
 \cdot& \cdot& \cdot& \cdot& \cdot& 1& \cdot & \cdot& \cdot  \\ \hline
 \cdot& \cdot& \cdot& \cdot& \cdot& \cdot & 1& \cdot& \cdot \\
 \cdot & \cdot& \lambda & \lambda & \cdot& \cdot& \cdot& \lambda & \cdot \\
 -1& \cdot& \cdot& \cdot& -1 & \cdot& \cdot& \cdot& 1
  \end{array} \right)\ ,
\end{equation}
and hence $\mbox{Tr}(W_\lambda \rho_\gamma) <0 $,  if
\begin{equation}\label{lambda}
 \lambda < \frac{1-\gamma^2}{2 + \gamma^{-2}}\ .
\end{equation}
Actually, the maximal value of $\lambda$ is attainable  for
$\gamma^* = \sqrt{(\sqrt{3}-1)/2} \approx 0.605$. Therefore,
taking as $\rho_0$ the state $\rho_{\gamma^*}$, one finds
$\lambda_{[W_0,P]} = (1-\gamma^{*2})/(2+\gamma^{*{-2}}) \approx
0.133$. This way it is shown that $W_\lambda$, with $0\leq \lambda
< \lambda_{[W_0,P]}$, defines an atomic EW. We may still modify
$W_\lambda$ by adding for example a positive operator
$Q=3|\varphi\>\<\varphi|$, where
\begin{equation}\label{}
    \varphi = \frac{1}{\sqrt{3}} ( e_1 \ot e_2 + e_2 \ot e_3 + e_3 \ot
    e_1) \ ,
\end{equation}
that is
\begin{equation}\label{}
    W_{\lambda,\mu} = W_0 + \lambda P + \mu Q\ .
\end{equation}
One finds the following matrix representation
\begin{equation}\label{W-lm}
 W_{\lambda,\mu}\ = \  \left( \begin{array}{ccc|ccc|ccc}
 1 &  \cdot& \cdot& \cdot& -1 & \cdot& \cdot& \cdot & -1 \\
 \cdot& 1+\mu &\cdot& \cdot& \cdot& \mu& \mu& \cdot& \cdot\\
 \cdot& \cdot & \lambda & \lambda & \cdot& \cdot& \cdot& \lambda & \cdot  \\ \hline
 \cdot& \cdot & \lambda & \lambda & \cdot& \cdot& \cdot& \lambda & \cdot \\
 -1 & \cdot& \cdot& \cdot& 1 & \cdot& \cdot& \cdot& -1 \\
 \cdot& \mu & \cdot& \cdot& \cdot& 1+\mu & \mu & \cdot& \cdot  \\ \hline
 \cdot& \mu & \cdot& \cdot& \cdot& \mu & \mu & \cdot& \cdot \\
 \cdot & \cdot& \lambda & \lambda & \cdot& \cdot& \cdot& \lambda & \cdot \\
 -1& \cdot& \cdot& \cdot& -1 & \cdot& \cdot& \cdot& 1
  \end{array} \right)\ .
\end{equation}
Now, $\mbox{Tr}(W_{\lambda,\mu} \rho_\gamma) <0 $,  if $\lambda$
satisfies (\ref{lambda}) and
\begin{equation}\label{}
 \mu < \frac{1-\gamma^2 - \lambda(2 + \gamma^{-2})}{2+\gamma^2}\ .
\end{equation}
Interestingly, applying our method to a pair $(W_0,\rho_\gamma)$
we constructed an atomic EW $W_{\lambda,\mu}$ which has a
circulant structure analyzed in \cite{CIRCULANT}. Therefore, it
may be used to test quantum entanglement within  a class of
circulant PPT states \cite{CIRCULANT} (see also \cite{PPT-nasza}).
To the best of our knowledge this is the first example of a
`circulant atomic' EW.

Actually, this example may be  generalized for $d \ot d$ case.
Consider the following set of Hermitian operators:
\begin{equation}\label{W_dk}
    W_{d,k} := \sum_{i,j=1}^d e_{ij} \ot X^{d,k}_{ij}\ ,
\end{equation}
where the $d \times d$ matrices $X^{d,k}_{ij}$ are defined as
follows:
\begin{equation}\label{}
X^{d,k}_{ij} = \left\{ \begin{array}{ll} (d-k-1)e_{ii} +
\sum_{l=1}^k
e_{i+l,i+l}  &,\   i=j \  \\
- e_{ij}  &, \  i\neq j \  \end{array} \right. \ .
\end{equation}
For $d=3$ and $k=1$ the above formula reconstructs $W_0$ defined
in (\ref{W0-C}). Again, $W_{d,k}$ are related via
Choi-Jamio{\l}kowski isomorphism to the family of positive maps
\cite{Ando}
\begin{equation}\label{}
    \tau_{d,k}(x) = (d-k) \varepsilon(x) + \sum_{l=1}^k
    \varepsilon(S^l x S^{* l}) - x \ , \ \ \ \ \ x \in M_d \ ,
\end{equation}
where $\varepsilon(x) = \sum_{i=1}^d x_{ii} e_{ii}$, and $S$ is
the shift operator defined by $Se_i = e_{i+1} \ (\mbox{mod}\ d)$.
The positivity of $\tau_{d,k}$ for $k=1,\ldots,d-1$ was shown by
\cite{Ando} (for $k=d-1$ this map is completely copositive) and
Osaka shown that $\tau_{d,1}$ is atomic. Finally, it was shown by
Ha \cite{Ha} that it is atomic for $k=1,\ldots,k-2$. Therefore, it
proves the atomicity of $W_{d,k}$. Ha's proof is based on the
construction of the family of states $\rho_\gamma \in \mathbf{V}_2
\cap \mathbf{V}^2$:
\begin{equation}\label{}
    \rho_\gamma = \frac{1}{N_\gamma}\, \sum_{i,j=1}^d e_{ij} \ot A^\gamma_{ij}\ ,
\end{equation}
where the $d \times d$ matrices $A^\gamma_{ij}$ are defined as
follows:
\begin{equation}\label{}
A^\gamma_{ij} = \left\{ \begin{array}{ll} e_{ij} & , \ i\neq j
\\ e_{11}  + a_\gamma e_{22} + \sum_{l=3}^{d-1}
e_{ll} + b_\gamma e_{dd} & ,\   i=j=1 \  \\
 S^{j-1}A_{11}S^{* j-1}  & , \  i= j\neq 1 \  \end{array} \right. \
 ,
\end{equation}
with
\begin{equation}\label{}
    a_\gamma = \frac{1}{d}(\gamma^2 + d-1)\ , \ \ \ \ b_\gamma = \frac{1}{d}(\gamma^{-2} +
    d-1)\ ,
\end{equation}
 and the normalization factor
\begin{equation}\label{}
    N_\gamma = d^2 - 2 + \gamma^2 + \gamma^{-2}\ ,
\end{equation}
which reproduces (\ref{N-3}) for $d=3$. One shows \cite{Ha} that
$\rho_\gamma \in \mathbf{V}_2 \cap \mathbf{V}^2$ and
$\mbox{Tr}(W_{d,k} \rho_\gamma) = (\gamma^2-1)/N_\gamma$. Hence,
for $\gamma<1$, the family of states $\rho_\gamma$ is detected by
each $W_{d,k}$ for $k=1,\ldots,d-2$. It is therefore clear that
any convex combination
\begin{equation}\label{}
  W_d[\mathbf{p}] :=  \sum_{k=1}^{d-2} p_k \, W_{d,k} \ , \ \ \ \
  \mathbf{p}=(p_1,\ldots,p_{d-2})\ ,
\end{equation}
the new EW $W_d[\mathbf{p}]$ is still atomic. Following
3-dimensional example one may easily construct out of a pair
$(W_{d,k},\rho_\gamma)$ a family of new EWs.

\section{Multipartite entanglement witnesses} \label{MULTI}

Let us note, that the above construction works perfectly for
multipartite case. Consider $N$-partite system living in
$\mathcal{H}= \mathcal{H}_1 \ot \ldots \ot \mathcal{H}_N$. A state
$\rho_0$ in $\mathcal{H}$ is entangled if there exists an
entanglement witness $W_0 \in \mathcal{B}(\mathcal{H}_1 \ot \ldots
\ot \mathcal{H}_N)$ such that:
\begin{enumerate}
\item $\mbox{Tr}(W_0 \sigma_{\rm sep}) \geq 0$ for all $N$-separable
states $\sigma_{\rm sep}$,

\item  $\mbox{Tr}(W_0 \rho_0) < 0$.
\end{enumerate}
In the multipartite case a set of PPT states may be generalized as
follows. For each binary $N$-vector
$\bsigma=(\sigma_1,\ldots,\sigma_N)$ one introduces a class of
$\bsigma$-PPT states: $\rho$ is  $\bsigma$-PPT iff
\begin{equation}\label{}
  \tau^{\sbsigma}\rho :=  (\tau^{\sigma_1} \ot \ldots \ot\tau^{\sigma_N})\,\rho \geq 0 \ .
\end{equation}
Finally, an entanglement witness $W$ is $\bsigma$-decomposable if
it may represented as the following sum
\begin{equation}\label{}
    W = Q_1 + \tau^{\sbsigma}Q_2 \ ,
\end{equation}
where $Q_1$ and $Q_2$ are positive operators in
$\mathcal{B}(\mathcal{H}_1 \ot \ldots \ot \mathcal{H}_N)$.
Clearly, $\bsigma$-decomposable EW cannot detect entangled
$\bsigma$-PPT state.

Suppose, that an entangled $N$-partite $\bsigma$-PPT state
$\rho_0$ is detected by $\bsigma$-indecomposable entanglement
witness $W_0$. Therefore, if $\sigma_{\rm sep}$ is an arbitrary
$N$-separable state, then the following convex combination
\begin{equation}\label{cc}
    \rho_\alpha = (1-\alpha)\rho_0 + \alpha \sigma_{\rm sep} \ ,
\end{equation}
defines $\bsigma$-PPT entanglement state   for any $0 \leq \alpha
< \alpha_{[\rho_0,\sigma_{\rm sep}]}$, with
\begin{equation}\label{}
\alpha_{[\rho_0,\sigma_{\rm sep}]} := \sup\, \{ \alpha \in [0,1]\
|\ \mbox{Tr}(W_0\rho_\alpha)< 0 \} \ .
\end{equation}
 This construction gives rise to an
open convex set
\begin{equation}\label{}
\mathcal{S}^{\rm PPT}_\sbsigma[W_0|\rho_0] := \Big\{ \rho_\alpha\
\Big| \ 0 \leq \alpha < \alpha_{[\rho_0,\sigma_{\rm sep}]}\ \ \& \
{\rm aribitrary}\ \ \sigma_{\rm sep} \ \Big\} \ .
\end{equation}
Similarly,  let $P$ be an arbitrary positive semidefinite operator
in $\mathcal{B}(\mathcal{H}_1 \ot \ldots \ot
 \mathcal{H}_N)$ and consider one-parameter family of operators
\begin{equation}\label{W-lambda}
    W_\lambda = W_0 + \lambda P\ , \ \ \ \ \ \lambda \geq 0 \ .
\end{equation}
Let us observe that for any $0 \leq \lambda < \lambda_{[W_0,P]}$
with
\begin{equation}\label{}
\lambda_{[W_0,P]} := \sup\, \{ \lambda \geq 0 \ |\
\mbox{Tr}(W_\lambda\rho_0)< 0 \} \ ,
\end{equation}
$W_\lambda$ defines $\bsigma$-indecomposable EW detecting the
state $\rho_0$. This construction gives rise to a dual open convex
set
\begin{equation}\label{}
\mathcal{W}^{\rm ind}_\sbsigma[W_0|\rho_0] := \Big\{ \ W_\lambda\
\Big| \ 0 \leq \lambda < \lambda_{[W_0,P]}\ \ \&\ \ {\rm
aribitrary }\ P\geq 0 \ \Big\} \ .
\end{equation}

\section{Conclusions}

A simple and general method for constructing indecomposable EWs
was presented. Knowing  one EW $W_0$ and the corresponding
entangled PPT state $\rho_0$ detected by $W_0$, one is able to
construct new EWs and new PPTES. In particular one may apply this
method to construct new examples of atomic EWs which are crucial
to distinguish between separable and entangled states. Moreover,
one may apply the same strategy to construct EWs for multipartite
systems.

What we can do if only one element from the above pair is
available?  Note, that a nonpositive Hermitian operator in
$\mathcal{B}(\mathcal{H}_1 \ot \mathcal{H}_2)$ may be always
written as a difference of two positive operators $P$ and $Q$:
\begin{equation}\label{QP}
    W = Q - P\ ,
\end{equation}
and, as is well know, most of known EWs have this form with $Q$
being separable (very often $Q \propto \mathbb{I}_1 \ot
\mathbb{I}_2$, but following \cite{TT} one may look for more
general form of $Q$) and $P$ being entangled (for example
maximally entangled pure state). Let $W$ defined in (\ref{QP}) be
an EW detecting an NPT (and hence entangled) state $P$. Is $W$
indecomposable? One may try to look for the states detectable by
$W$ in the following form
\begin{equation}\label{}
    \rho_\alpha = (1-\alpha) P + \alpha\sigma_{\rm sep}\ ,
\end{equation}
where $\sigma_{\rm \alpha}$ is a separable state. Now, mixing an
NPT state $P$ with $\sigma_{\rm sep}$ may result in a PPT state.
Hence, if $\rho_\alpha$ becomes  PPT for some $\alpha>0$, and it
is still detected by $W$, then $W$ is necessarily indecomposable
EW.

Conversely, given a PPTES state $\rho$ one may try to construct
the corresponding (indecomposable) EW detecting $\rho$. This
problem is in general very hard since it is extremely difficult to
check weather $W$ satisfies $\mbox{Tr}(W\sigma_{\rm sep})\geq 0$
for {\it all} separable $\sigma_{\rm sep}$. One example of such
construction is provided via unextendible product bases by Terhal
\cite{Terhal2}.

It is clear, that presented method provides new classes of
indecomposable (and atomic) linear positive maps (for recent
analysis of atomic maps see \cite{OSID-W}). In particular a
positive map corresponding to  $W_{\lambda,\mu}$ defined in
(\ref{W-lm}) provides a considerable generalization of the Choi
map. On may try to look for other well know positive
indecomposable maps and to perform `deformation' within the class
of indecomposable maps. Any new examples of such maps provide
important tool for the  studies of quantum entanglement.

\section*{Acknowledgement} This work was partially supported by the
Polish Ministry of Science and Higher Education Grant No
3004/B/H03/2007/33.

\end{document}